\documentclass[12pt]{article}

\usepackage{epsf}
\usepackage{amsmath,amssymb}
\usepackage{latexsym}
\usepackage{graphicx,color}
\usepackage{wrapfig}
\usepackage{latexsym}
\usepackage{graphics}

\usepackage{color}
\usepackage{delarray,color}
\usepackage{fancybox}  

\newcommand {\beq}{\begin{eqnarray}}
\newcommand {\eeq}{\end{eqnarray}}

\newcommand{\be}{\begin{equation}}
\newcommand{\ba}{\begin{eqnarray}}
\newcommand{\ea}{\end{eqnarray}}
\newcommand{\ee}{\end{equation}}

\newcommand{\la}{\langle}
\newcommand{\lb}{\rangle}

\newcommand{\beqa}{\begin{eqnarray}}
\newcommand{\eeqa}{\end{eqnarray}}

\newcommand{\unit}{\hbox to 3.8pt{\hskip1.3pt \vrule height 7.4pt
    width .4pt \hskip.7pt \vrule height 7.85pt width .4pt \kern-2.4pt
    \hrulefill \kern-3pt \raise 3.7pt\hbox{\char'40}}}

\def\matt[#1,#2,#3,#4]{\left(%
\begin{array}{cc} #1 & #2 \\ #3 & #4 \end{array} \right)}

\usepackage{tabularx}

\catcode`\@=11
\@addtoreset{equation}{section}

\catcode`@=12
\relax

\newcommand{\tr}{\mathrm{Tr}}

\begin{document}

\begin{titlepage}

\setcounter{page}{0}

\renewcommand{\thefootnote}{\fnsymbol{footnote}}

\begin{flushright}
YITP-15-77 \\
\end{flushright}

\vskip 1.35cm

\begin{center}
{\Large \bf 
A Nonperturbative Proof 
of Dijkgraaf-Vafa Conjecture
}

\vskip 1.2cm 

{\normalsize
Seiji Terashima\footnote{terasima(at)yukawa.kyoto-u.ac.jp}
}

\vskip 0.8cm

{ \it
Yukawa Institute for Theoretical Physics, Kyoto University, Kyoto 606-8502, Japan
}

\end{center}

\vspace{12mm}

\centerline{{\bf Abstract}}

In this note 
we exactly compute the gaugino condensation 
of an arbitrary four dimensional ${\cal N}=1$ 
supersymmetric gauge theory in confining phase,
using the localization technique.
This result gives a nonperturbative proof of the 
Dijkgraaf-Vafa conjecture.

\end{titlepage}
\newpage

\tableofcontents
\vskip 1.2cm 

\section{Introduction and summary}

Analytic computations in quantum field theories
are important, but very hard, in general.
Important quantum field theories
in which we can compute some quantities exactly are
supersymmetric (SUSY) field theories.
The localization technique for SUSY field theories, 
originated in \cite{Witten1} \cite{Witten2}, is a general way 
to compute them exactly.
Recently, this technique is applied to 
various kinds of SUSY field theories (for examples, \cite{Kap}-\cite{Hashimoto}), 
after the important work by Pestun \cite{Pestun}.
It should be stressed that
using the localization technique we can compute
non-topological quantities.
In particular, using it, we can compute
the gaugino condensation in four dimensional ${\cal N}=1$ SUSY
Yang-Mills theories
in confining phase \cite{ST}.\footnote{
The gaugino condensation were computed various ways, see \cite{g1}-\cite{g2}.}

In this paper,
we will compute the gaugino condensation of four dimensional ${\cal N}=1$ SUSY
gauge theories with general chiral multiplets and a superpotential 
in confining phase.
In order to do this, we first integrate out the 
chiral multiplets, while keeping the vector multiplets.
This is consistent because we can deform the theory without changing
the gaugino condensation (i.e. using the localization technique)
such that the theory is arbitrary weak coupling
\cite{ST}.\footnote{
Here, we consider the theory on ${\mathbf R}^3 \times S^1_{R}$ and
then taking the $R \rightarrow \infty $ limit.
(The gaugino condensation does not depend on $R$.)
With the non-trivial v.e.v. of the Wilson line around $S^1_R$,
the symmetry breaking will occur at the very high scale compared with
the scale $\tilde{\Lambda}$ which is the effective dynamical scale
determined by the deformed action.
Thus, the computations reduced to 3d Abelian theory
in the low energy region much below the $1/R$.
Withtout this infra-red cut-off $R$, the deformed action will not be 
weak coupling because of the low energy modes below the
$\tilde{\Lambda}$ which remain strong coupling.
}
This integration of the chiral multiplets
can be done perturbatively and we only need the 
effective superpotential.
Thus, this can be done by the methods used in \cite{Zanon} and \cite{CDSW}.

After the integration of the chiral multiplets,
we have ${\cal N}=1$ SUSY gauge theories with only vector multiplets
and a superpotential.
As shown in \cite{CDSW}, the superpotential is a function of
the gaugino bi-linear $S$ only (and the coupling constants in the
original superpotential).
For this theory, we will compute the gaugino condensation.
Therefore, we can compute the 
gaugino condensation in four dimensional ${\cal N}=1$ SUSY
gauge theories with general chiral multiplets and a superpotential.

The Dijkgraaf-Vafa conjecture is that
the glueball superpotential of
the ${\cal N}=2$ SUSY $U(N)$ gauge theory deformed 
by a superpotential is computed by a corresponding matrix model \cite{DV}.
There are ``proofs'' of this conjecture, i.e. \cite{Zanon} and \cite{CDSW},
however, both in \cite{Zanon} and \cite{CDSW} perturbative integrations 
of chiral multiplets were computed and the nonperturbative dynamics of 
the gauge fields were (implicitly) assumed to be just adding
the Veneziano-Yankielowicz superpotential to the non-trivial
superpotential
which was obtained by the integration of the chiral multiplets.\footnote{
More precisely, in \cite{CDSW} using the generalized Konishi anomaly 
equation to the 1PI effective action written by $S$ we can justify 
the addition of the Veneziano-Yankielowicz superpotential.
However, as stressed in \cite{CDSW} this only works for the case without
symmetry breaking because there are no coupling constants to $S_i$ where
$S=\sum_i S_i$. We thank Y. Nakayama for the useful discussions 
on this point.
In \cite{CDSW} it was also noted that the generalized
Konishi anomaly would have the higher loop corrections.}
In this paper, we show that
the perturbative superpotential with the Veneziano-Yankielowicz superpotential indeed 
gives the correct gaugino condensation.
This can be regarded as a nonperturbative proof of the 
Dijkgraaf-Vafa conjecture.\footnote{
In \cite{Fucito},
the gaugino condensation was computed in the way, which
is different from ours and is related to the
${\cal N}=2$ Seiberg-Witten theory.
The close connection of the gaugino condensation to 
Seiberg-Witten theory was also discussed in   
\cite{Matone}.
In \cite{Ferrari}, an off shell extension of the 
vacuum expectation value was used to compute the 
gaugino condensation and it was claimed to give 
a non-perturbative proof of the Dijkgraaf-Vafa conjecture.
}

It should be noted that
we can compute the gaugino condensation 
for any four dimensional ${\cal N}=1$ SUSY gauge theories 
(with a Lagrangian and in confining phase)
according to the discussions in this paper.\footnote{
The perturbative superpotential should be computed using the results in
\cite{Zanon} and \cite{CDSW} for general ${\cal N}=1$ gauge theories,
for examples, \cite{Oz} \cite{Nakayama},
however, it would be difficult to have the superpotential
in an explicit closed form.}
It would be interesting to study applications of our method.
We hope to return this problems in future.

The organization of this paper is as follows:
In section 2 we compute the 
gaugino condensation for four dimensional ${\cal N}=1$ SUSY gauge theory 
with only vector multiplets and a generic action.
In section 3 we show that our results imply
the nonperturbative proof of the Dijkgraaf-Vafa conjecture.

\section{Gaugino condensation in theory with a generic superpotential}

In this section, we will consider four dimensional  ${\cal N}=1$ SUSY gauge theory 
with vector multiplets only (no chiral multiplets) on ${\mathbf R}^3 \times S^1_R$ with 
a simple gauge group $G$ and the following superpotential:
\begin{eqnarray}
 W_{V}(\tau_0,g_i) = 
2 \pi i \tau_0 S+F(S,g_i),
\label{WDV}
\end{eqnarray}
where $\tau_0$ and $g_i$ are complex constants,
$S(y,\theta)=S_0(y)+\theta S_1(y)+\theta \theta S_2(y)$ is the 
glueball superfield whose lowest component is  
the gaugino bilinear
$S_0 \sim \tr (\lambda \lambda) $ 
and $F(S,g_i)$ is a function of $S$ and 
coupling constants $g_i$.
Here we do not assume the K\"ahler potential is canonical.
Note that terms containing 
$\tr ( (\lambda \lambda)^n )  $ with $n>1$ and 
terms with derivatives 
are regarded as the K\"ahler potential \cite{CDSW}.
Thus, this superpotential represents a general superpotential
for a theory without chiral multiplets.

Note that the polynomials of $S$ in the superpotential
are composite operators and should be defined with a regularization,
for example, a point splitting.
Thus, the classical constraints are not imposed on these composites.

We will compute the effective superpotential 
for this theory and determine the vacua and then
compute the gaugino condensation 
\begin{eqnarray}
 \bar{S} \equiv \langle S \rangle.
\end{eqnarray}
Using the result of \cite{ST}, 
we can compute them in the weak Yang-Mills coupling constant
by the localization technique.
In a superspace, this is realized by adding 
\begin{eqnarray}
 t  \int d^4x d^2 \bar{\theta} \,
\bar{W}_{\dot{\alpha}}  \bar{W}^{\dot{\alpha}},
\label{regpot}
\end{eqnarray}
to the anti-superpotential
with $t \gg \infty$.
The dynamical scale $\tilde{\Lambda}$ of 
the theory with the additional term (\ref{regpot})
can be arbitrary low.
With this deformed action,
what we need to do is only
semi-classical computations around the anti-self dual (ASD) connections.
Note that the radius $R$ of $S^1_R$ plays as 
a infra-red regulator, which makes the deformed theory indeed weak coupling
\cite{ST}.

As in \cite{ST},
in order to determine the vacua and evaluate the gaugino condensation,
we would like to compute the expectation value of the gaugino bi-linear:
\begin{eqnarray}
X
\equiv \la \lambda(x-b) \lambda(x) 
\lb
= \la \lambda(x-b) \lambda(x) 
e^{\int d^4 y d^2 \theta F(S(y,\theta),g_i)}
\lb_0
=
 \la \lambda(x-b) \lambda(x) 
e^{\int d^4 y \delta_1 \delta_2 F(S_0(y),g_i)}
\lb_0,
\end{eqnarray}
where 
$\la \cdots \lb_0$ means the expectation value 
with the gauge coupling $\tau_0$ and a K\"ahler potential
without the superpotential $F$.
We also used $\delta_1,\delta_2$ which are SUSY transformations,
corresponding to $\theta_1, \theta_2$.
In order to evaluate this, it seems to have to consider 
ASD connections with many fermion zeromodes
because the superpotential term $F$ includes the fermions.
To do this explicitly is interesting, however,
in this paper we will use a different method.

First, 
we introduce $\tilde{S}(x)$ as a
constant shift $\bar{S}$ of $S(x)$:
\begin{eqnarray}
\tilde{S}(x) \equiv S(x)-\bar{S},
\end{eqnarray}
and define $G$ from $F$ 
by subtracting the zeroth and linear term in $\tilde{S}$:
\begin{eqnarray}
 G(\tilde{S}(x),g_i,\bar{S}) \equiv
 F(\bar{S}+\tilde{S}(x),g_i) -
\left(  F(S) \big|_{S=\bar{S}} +
\frac{d F(S)}{d S} \big|_{S=\bar{S}} \,\, \tilde{S}(x)
\right).
\end{eqnarray}
Then, in terms of $\tilde{S}_0(x)$,
we can express $X$ as
\begin{eqnarray}
X= \la \lambda(x-b) \lambda(x) 
e^{\int d^4 y \delta_1 \delta_2 G(\tilde{S}_0(y),g_i,\bar{S}_0)}
\lb_{\bar{S}},
\label{x1}
\end{eqnarray}
where 
\begin{eqnarray}
 \la (\cdots) \lb_{\bar{S}}
\equiv \la (\cdots) \
e^{
\int d^4 y 
\delta_1 \delta_2 
\left(
F(S)|_{S=\bar{S}} +\,
\frac{d F(S)}{d S} \big|_{S=\bar{S}_0} \,
\tilde{S}_0(y) 
\right)}
\lb_0,
\label{sbar}
\end{eqnarray}
where we left $F(S)|_{S=\bar{S}}$ term,
which vanishes for constant $\bar{S}$,
for later convenience.
This term will be relevant 
if we regard $\bar{S}$ as a background constant chiral superfield.
Note that the action of $ \la (\cdots) \lb_{\bar{S}}$ 
contains only linear term in $\tilde{S}$.

Now we take the constant $\bar{S}$ to satisfy 
\begin{eqnarray}
 \bar{S}=\la S (x)\lb_{\bar{S}},
\label{dets}
\end{eqnarray}
which means $\la \tilde{S}(x) \lb_{\bar{S}}=0$.
We will see later that this $\bar{S}$ indeed gives the 
gaugino condensation, i.e. $ \bar{S}=\la S (x)\lb$.
The condition (\ref{dets}) will be 
a self-consistent equation.\footnote{
If we do not regard $\tau_0$ as a chiral superfield,
we have $\langle S(x) \rangle=\langle S_0(x) \rangle$,
i.e. $\langle S_1(x) \rangle=\langle S_2(x) \rangle=0$,
because $S_1$ and $S_2$ are not Lorentz invariant. }
Then we expand
$e^{\int d^4 y \delta_1 \delta_2 G(\tilde{S}_0(y),g_i,\bar{S}_0)}$
in (\ref{x1}) in terms of $\tilde{S_0}$.
It will be a linear combination of $1$ and
\begin{eqnarray}
 I_n= \int d^4 x \delta_1 \delta_2 ( C_n \, (S_0(x))^n),
\end{eqnarray}
where $C_n=C_n(\bar{S}_0,g_i)$ is determined by $G$. 
We can easliy see that $I_n$ satisfies $\bar{\delta}_i I_n=0$,
because $\bar{\delta}_i S_0(x)=0$ and
$[ \bar{\delta}_i, \delta_1 \delta_2 ]$ is a space derivative.
Here $\bar{\delta}_i$ is the SUSY transformation 
corresponding to $\bar{\theta}_i$.
Furthermore, we can show
\begin{eqnarray}
 I_n= \int d^4 x 
\left(
\delta_1 \delta_2 ( 
C_n \, (S_0(x))^{n-1}
e^{i a^\mu \partial_\mu} S_0(x)  \, )
+ \bar{\delta}_1 \delta_1 \delta_2 (\cdots)
+ \bar{\delta}_2 \delta_1 \delta_2 (\cdots)
\right).
\end{eqnarray}
This follows from
$\frac{\partial}{\partial a_\nu}  e^{i a^\mu \partial_\mu} S_0(x)=
i (\bar{\delta} \sigma^\mu \delta) e^{i a^\mu \partial_\mu} S_0(x)$
which means 
$e^{i a^\mu \partial_\mu} S_0(x)=S_0(x)+
 (\bar{\delta} \sigma^\mu \delta) (\cdots)$.
Therefore, for the $\bar{\delta}_i$-closed correlators,
which we are considering,
we can replace 
\begin{eqnarray}
I_n \rightarrow \int d^4 x \delta_1 \delta_2 
\left(
 C_n \prod_{j=1}^n  S_0(x+a_j)
\right), 
\end{eqnarray}
where $a_j$ is an arbitrary constant.

Note that we can do this replacement 
for each $I_n$ in a product of $I_n$s in the expansion 
of the exponential with 
different $a_j$ for each $I_n$.
Then, $X$ will be written by a linear combination of the following form:
\begin{eqnarray}
 \la
\lambda(x-b) \lambda(x) 
\prod_{\alpha=1}^M 
\left(
\int d^4 x^\alpha 
\delta_1 \delta_2 
\left( C_n 
\prod_{i=1}^{m_\alpha} 
\tilde{S}_0(x^\alpha + a_i^\alpha)
\right)
\right)
\lb_{\bar{S}},
\label{iso} 
\end{eqnarray}
where $m_\alpha \geq 2$.
Now we will take a following ``large separations'' limit:
$|a_j^\alpha| \rightarrow \infty$
with $|a_j^\alpha -a_k^\beta| \rightarrow \infty$ $((j,\alpha) \neq (k,\beta))$
and $|a_j^\alpha-b| \rightarrow \infty$.
Then, we will use the clustering properties to
factorize the correlator for each 
$\la \tilde{S}(x^\alpha+a_i^\alpha)\lb$ if $|x^\alpha+a_i^\alpha|$
is not close to any other insertion points.
Here we can see that
$\la \tilde{S}_0(x^\alpha+a^\alpha_i) \lb_{\bar{S}}=0$,
$\la \delta_i \tilde{S}_0(x^\alpha+a^\alpha_i) \lb_{\bar{S}}=0$
and 
$\la \delta_1 \delta_2 \tilde{S}_0(x^\alpha+a^\alpha_i) \lb_{\bar{S}}=0$
by the definition
of $\tilde{S}$, i.e.  $\la \tilde{S}(x) \lb_{\bar{S}}=0$.
Furthermore, 
the number of the points of the integration is $M$,
which is strictly smaller than the number of $a_i^\alpha$
because $m_\alpha \geq 2$.
Therefore, there is at least an isolated insertion of an operator
which makes (\ref{iso}) vanishes and we find 
\begin{eqnarray}
X= \la \lambda(x-b) \lambda(x) 
e^{\int d^4 y d \theta^2 F(S(y,\theta),g_i)}
\lb_{0}
=\la \lambda(x-b) \lambda(x) \lb_{\bar{S}},
\end{eqnarray}
{}from which we can evaluate the superpotential, the vacua
and the gaugino condensation as in \cite{Davies, ST}.
Note that this also implies $\la S_0(x) \lb =\la S_0(x) \lb_{\bar{S}}$.

As we can see from (\ref{sbar}),
the $\la \cdots \lb_{\bar{S}}$ is just replacing the 
coupling constant $\tau_0$ to
\begin{eqnarray}
 \tilde{\tau}= \tau_0+\frac{1}{2 \pi i} 
\frac{\partial F(S)}{\partial S} \big|_{S=\bar{S}}.
\label{tandt0}
\end{eqnarray}
Thus, the superpotential and the vacua\footnote{
In order to find the vacua, 
we need to evaluate 
$\partial W_{eff}(X,\tau_0)/\partial X=0$, 
where $X$ is the (would-be) moduli.
With this and 
$ \partial ( W_{eff}(X,\tau_0) -f(\bar{S}) ) /\partial \tau_0=\bar{S}(X,\tau_0)$,
we see that 
\begin{eqnarray}
 \partial \bar{S}(X,\tau_0)/\partial X=0,
\end{eqnarray}
is a solution.
Thus, we can think that $\bar{S}$ does not depend on $X$.
}
are
found as in \cite{Davies, DHK, ST} by 
the semi-classical computations around the fundamental monopoles
which have two fermion zeromodes.
Note that the 1-loop factor in the localization technique 
only contributes to the K\"ahler potential \cite{Poppitz}.
More precisely,
the definition of $\la (\cdots) \lb_{\bar{S}}$, (\ref{sbar}), 
is the path-integral with the superpotential
\begin{eqnarray}
\tilde{W}=f(\bar{S})
+2 \pi i \, \tilde{\tau} S,
\label{sbar2a}
\end{eqnarray}
where 
\begin{eqnarray}
 f(\bar{S})=(F(S)-S \frac{\partial F(S)}{\partial S})|_{S=\bar{S}}.
\end{eqnarray}
Thus,
the effective superpotential is 
\begin{eqnarray}
 W_{eff}= c_2 \, \omega \, e(G) \, \tilde{\Lambda}^3 + f(\bar{S}),
\end{eqnarray}
where $c_2$ is the dual coxeter number of $G$ and 
\begin{eqnarray}
 e(G)=\prod_{j=0}^{r} (k_j^* \alpha_j^*/2 )^{-k_j^*/2}  ,
\,\,\,\, \omega^{c_2}=1,
\end{eqnarray}
for example, $e(SU(N_c))=1$,
and the $\tilde{\Lambda}$ is the dynamical scale 
in the 1-loop Pauli-Villars regularization which is defined by
\begin{eqnarray}
 \tilde{\Lambda}^3=\mu^3 \frac{1}{g^2(\mu)}
 \exp\frac{2 \pi \tilde{\tau} (\mu)}{c_2},
\end{eqnarray}
where $g^2(\mu)$ comes from the path-integral measure 
which is defined with the coupling constant $\tau_0$.

Now we will consider the gaugino condensation $\la S \lb(=\bar{S})$.
We have seen that the original superpotential $W_{V}$ can be written as
\begin{eqnarray}
 W_{V}=f(\bar{S}) +2 \pi i \tilde{\tau} S 
+ G(\tilde{S},g_i,\bar{S}),
\end{eqnarray}
then, the effective potential should give 
\begin{eqnarray}
\frac{\partial}{ \partial \tilde{\tau}}
 W_{eff} (\tilde{\tau},\bar{S}(\tilde{\tau}))
=\la 2 \pi i S + \frac{\partial}{ \partial \tilde{\tau}} f(\bar{S})
+\frac{\partial}{ \partial \tilde{\tau}} G
\lb
=
\la 2 \pi i S + \frac{\partial}{ \partial \tilde{\tau}} f(\bar{S}) \lb,
\end{eqnarray}
where we have used $\la \frac{\partial}{ \partial \tilde{\tau}} G \lb=0$
which follows from $\la (\mbox{polynomials of } \tilde{S}) \lb=0$.
Thus, the gaugino condensation $\bar{S}$ can be computed 
using
\begin{eqnarray}
 \frac{\partial}{ \partial \tilde{\tau}}\left(
 W_{eff} (\tilde{\tau},\bar{S}(\tilde{\tau}))
-  f(\bar{S}) \right)
=\la 2 \pi i S 
\lb.
\end{eqnarray}

The result is
\begin{eqnarray}
 \bar{S}=\la S(x) \lb = e(G) w \tilde{\Lambda}^3.
\label{sbar2}
\end{eqnarray}
The superpotential is evaluated to 
\begin{eqnarray}
 W_{eff}= c_2 \, \bar{S} + f(\bar{S}),
\end{eqnarray}
where $\bar{S}$ is determined by (\ref{sbar2}).
We will also define
a dynamical scale $\Lambda$ 
in the 1-loop Pauli-Villars regularization of the coupling 
constant $\tau_0$,
\begin{eqnarray}
 \Lambda^3=\mu^3 \frac{1}{g^2(\mu)}
 \exp\frac{2 \pi \tau_0 (\mu)}{c_2}.
\end{eqnarray}
The relation between $\Lambda$ and $\tilde{\Lambda}$
is given by
\begin{eqnarray}
\tilde{\Lambda}^3
= \Lambda^3 
e^{\frac{1}{c_2}\frac{\partial F(\bar{S})}{\partial \bar{S}} },
\label{sbar3}
\end{eqnarray}
where we have used (\ref{tandt0})
and $\bar{S}=\bar{S}(\tilde{\Lambda})$ was given by (\ref{sbar2}).

Now we see that the following glueball superpotential 
reproduces the gaugino condensation and the effective superpotential:
\begin{eqnarray}
W_S (S,\Lambda)= -c_2 \, 
S \left( \ln \frac{S}{e(G) \Lambda^3} -1 \right)+F(S), 
\label{VY}
\end{eqnarray}
where we can think $W_S(S,\Lambda)$ as a function of $S$ 
and $\tilde{\Lambda}$ by using 
the relation $\tilde{\Lambda}= \tilde{\Lambda}(\Lambda)$.
Indeed, we find
\begin{eqnarray}
 \frac{\partial W_S}{\partial  S}= -c_2 \, \ln 
\frac{S}{e(G) \Lambda^3}+ \frac{\partial F(S)}{\partial S}=0,
\label{dw} 
\end{eqnarray} 
which is equivalent to 
$S= e(G) w \Lambda^3 
e^{\frac{1}{c_2}\frac{\partial F(S)}{\partial S} }$.
The superpotential $W_S$ is evaluated with (\ref{dw}) to
$W_S \rightarrow c_2 S +F(S)- S \frac{\partial F(S)}{\partial S}$,
which is the correct one.
Therefore, the glueball superpotential 
is the (\ref{VY}) which is just a sum of 
the Veneziano-Yankielowicz superpotential and the $F(S)$.

We can easily generalize the results to the theory with a semi-simple gauge group.
Let us consider a 4d ${\cal N}=1$ SUSY gauge theory of only vector
multiplets with
a gauge group $G=\otimes_a G_a$ is semi-simple
and a superpotential 
\begin{eqnarray}
 W_{V}(\tau_0^a,g_i) = 
\sum_{a=1}
\left(
2 \pi i \tau_0^a S_a +F(S_a,g_i)
\right).
\label{WDV2}
\end{eqnarray}
Following the previous discussions, 
we can easily see that
\begin{eqnarray}
 W_S (S_a,\Lambda)= 
-c_2^a \, \sum_a S_a \left( \ln \frac{S_a}{e(G_a) \Lambda_a^3} -1
\right)+F(S_a),
\end{eqnarray}
 which is evaluated to
$W_S \rightarrow c_2^a S_a +F(S_a)
- \sum_a S_a \frac{\partial F(S_a)}{\partial S_a}$
with
$S_a= e(G_a) w \Lambda_a^3 
e^{\frac{1}{c_2^a}\frac{\partial F(S_a)}{\partial S_a} }$.
Here, for $U(1)$ gauge group, there is no dynamically generated superpotential
and $S_a=0$.

\section{A proof of Dijkgraaf-Vafa conjecture}

Let us consider 4d ${\cal N}=1$  gauge theory
with gauge group $G$ and 
chiral multiplets couple to $G$.
With a generic tree level superpotential
\begin{eqnarray}
 W_{tree}=W_{tree}(g_i,\Phi_a),
\end{eqnarray}
where $g_i$ is the coupling constants\footnote{
If the low energy theory is a non-trivial conformal fixed point,
we will add an arbitrary small perturbation to the coupling constants or
a small deformation of the vacuum we choose.}
and $\Phi_a$ is the chiral superfields,
the theory is expected to be in a confining phase\footnote{
With the chiral multiplets, 
the Wilson loop will not behave the are law.
Thus, precisely speaking, 
the phase 
will not be a confining phase, but a phase with a mass gap
with possible free $U(1)$ factors.
For simplicity,
we will call it confining phase.}
and gaugino condensation is non-trivial,
which we will compute. 

We will compute the correlation functions 
of the operators insertions which satisfy 
${\bar{\delta}_i (\cal O)}=0$.
Thus, we can add the regularization term of  
the localization for the vector multiplets (\ref{regpot}) \cite{ST}.
Then, the theory is effectively in weak coupling 
and the effective dynamical scale can be set to
arbitrary low.

We can also add a large kinetic terms for the chiral multiplets.
\footnote{
The kinetic terms for the chiral superfields 
are written by the K\"ahler potential. The regularization term
(\ref{regpot}) for the vector
multiplet is the anti-holomorphic superpotential. Both of them
do not affect the effective superpotential and 
the correlation function of the operators in the chiral
rings.}
Then, we can integrate out the chiral multiplets perturbatively,
where the vector multiplets are regarded as a background
because the effective gauge coupling constant is very small 
by taking $t \rightarrow \infty$. 
Here we expand the bosonic fields in the chiral multiplets 
around the classical vacua.
Note that the 1-loop computaiton is exact in the usual localization
technique where 
we take the $t \rightarrow \infty$ limit with the 
regularization term $t \bar{\delta} V$ and rescaling of the fields.
In our case, the kinetic terms of the chiral multiplets 
contain the vector multiplets which is regarded as background fields.
Thus the saddle points of the large kinetic terms 
are non-trivial and integrations 
over the saddle points with the superpotential give a non-trivial
effective superpotential.
It will be interesting exactly follow this line and find the effective
superpotential
which should be a matrix model computation because the saddle points 
are essentially the zero modes of the chiral multiplets.

On the other hand, 
in \cite{Zanon}  
the perturbative computation
of the chiral multiplets with the vector multiplet background
was done by deforming the anti-superpotential appropriate way.
Furthermore, 
in \cite{CDSW}
it was shown that
the effective superpotential obtained by 
integrating out the chiral superfields
can be determined
by the generalized Konishi anomaly.
Thus, in this paper, we assume that 
the integrating out the chiral multiplets
is done by those methods.

Here, the chiral multiplets with classical superpotential
can have a non-trivial moduli space of vacua.
We have discrete set of vacua with a generic superpotential,
although, the moduli space need not to be discrete.
Here we assume the moduli space is discrete
by giving a small deformation of superpotential, for example 
a mass term, if it is needed.
Then, we redefine the chiral superfields
as $\Phi'_a=\Phi_a-\bar{\Phi}_a$ where $\bar{\Phi}_a$
is the value of $\Phi_a$ at the classical vacuum we have chosen.
The perturbative calculation is done around this.

Depending on the choice of the classical vacuum,
the original gauge group $G$ will be broken to a semi-simple gauge
group with $U(1)$ factors, which we will denote $G'$.
The glueball superfields $S_a$ are possible to be defined 
for this setting because the gauge symmetry is broken at 
very high energy scale compared to the effective dynamical scale of the gauge
theory which is lowered by the regulator term.\footnote{
The $S_a$ could be interpreted as a composite operator, 
like $\tr (\Phi^n W_\alpha W^\alpha )$ where $\Phi$ is the chiral
superfield, although we will not study this in this paper.
Note that such relations are just for the expectation values.
There is no effective superpotential fot $S_a$ for the original theory
as stressed in \cite{CDSW} because corresponding coupling constants 
are absent. }
In terms of $S_a$, we have the effective superpotential $W_{V}$, (\ref{WDV2}),
with $\tau_0^a=\tau_0$ for any $a$
after integration of the chiral superfields.

Then, applying the discussion in the previous section to
the effective action (\ref{WDV2}),
we conclude that the effective superpotential from which 
we can compute the gaugino condensation $\la S \lb=\sum_a \la S_a \lb$
is given by just adding the Veneziano-Yankielowicz superpotentials
for all simple gauge groups in $G'$ to
the effective action (\ref{WDV2}).

In particular, 
if we consider a chiral multiplet of the adjoint representation 
of $G$, 
it was shown in \cite{Zanon} and \cite{CDSW} 
that 
the perturbative effective superpotential for the chiral multiplet 
is equivalent to the one of the matrix
model of the Dijkgraaf and Vafa.
Then, 
the path-integral of the vector multiplets  
gives just the Veneziano-Yankielowicz terms
according to the discussion in the previous section.
The final effective action (\ref{WDV2}) is the one
conjectured in \cite{DV} .
Therefore, this can be regarded as a nonperturbative proof of the 
Dijkgraaf-Vafa conjecture.\footnote{
For the unbroken gauge group case, i.e. $G=G'$,
Dijkgraaf-Vafa conjecture was proved by using the 
generalized Konishi anomaly equation of the full effective action
\cite{CDSW} and taking the decoupling limit to get the
Veneziano-Yankielowicz terms.
}


\section*{Acknowledgments}

S.T. would like to thank 
Masato Taki
for his collaboration at the early stage of
this project and useful dicussions
and
Yu Nakayama for important comments and discussions. 
S.T. would like to thank  
K. Lee, K. Ohta, S. Rey, N. Sakai and P. Yi also for 
useful discussions.

\vspace{1cm}




\end{document}